\documentstyle[12pt]{article}
\begin{document}

\bibliographystyle{plain}
\author{\footnotesize Jong B. Choi\\
 \footnotesize {Department of Physics Education, Chonbuk National University,}\\
 \footnotesize {Chonju 561-756, Korea}\\
\footnotesize {Byeong S. Choi}\\
\footnotesize{Samrye Girl's High School, Wanju}\\
\footnotesize{Su K. Lee}\\
\footnotesize{Department of Physics, Chonbuk National University, Chonju 561-756, Korea}}
\title{Angular Ordering in Gluon Radiation}
\maketitle
\begin{abstract}
The assumption of angular ordering in gluon radiation is essential
to obtain quantitative results concerning gluonic behaviors. In order to
prove the validity of this assumption, we have applied our momentum
space flux-tube formalism to check out the angular dependences of
gluon radiation. We have calculated the probability amplitudes to get
new gluon, and have found that the new gluon is generally expected
to have the maximum amplitude when it is produced between the
momentum directions of the last two partons.
\end{abstract}
\newpage
%\section{}
 The gluonic behaviors in non-perturbative regions are not so much
understood that many phenomenological models are introduced to account for
the relevant processes such as small-x physics\cite{physics}, fragmentations into
hadrons\cite{hadrons}, and the confining aspects in bound states\cite{states}. In order to
provide for a systematic approach to these problems\cite{problems}, one of us has
developped flux-tube formalism\cite{formalism} in which gluonic flux-tubes are classified
and related to form topological spaces. General measures on the defined
topological spaces can be used to predict the gluonic structures of
hadrons\cite{hadron} and, when applied in momentum space, the particle multiplicity
distributions in jets can be analyzed systematically\cite{sys}. In this paper, we will
try to apply the developped flux-tube formalism to the prediction of gluon
radiation and to prove the validity of angular ordering assumption which
is extensively used in predicting particle fragmentations.
 
  It is well-known that the gluonic behaviors in perturbative region
can be described by CCFM equation\cite{equ}, which becomes A-P equation\cite{equa} for
large x and BFKL equation\cite{equat} for small x. The probability to have a
new gluon is usually given by\cite{by}
\begin{equation}
\rm{dp = \Delta_s \tilde{P} dz \frac{dq^{2}_{T}} {q^{2}_{T}} \Theta(\theta - \theta')} ,
\end{equation}
where $\Delta_s$ is the Sudakov form factor\cite{factor}, and $\tilde{\rm{P}}$ is the gluon-gluon
splitting function\cite{function}. The factor $\Theta(\theta - \theta')$ represents the condition that
the radiated gluons are ordered in angles. However, the final direction
of radiation cannot be determined by this short range condition. In order
to prove the validity of this assumption in long range scale\cite{scale}, we now
turn to a brief introduction of our flux-tube formalism.

 The starting point for a systematic description of flux-tubes rests
on the classification of flux-tubes, which are taken to start from
quark boundaries and to end at antiquark boundaries. It is sufficient
to count the number of quarks and antiquarks to classfy flux-tubes,
and we can represent the set of flux-tubes with $a$ quarks and $b$
antiquarks sitting at boundaries as $\rm{F_{a, \bar{b}}}$. Omitting the number 0,
except for $\rm{F_0}$ representing glueballs, mesonic flux-tubes are represented
by  $\rm{F_{1, \bar{1}}}$, and baryonic and antibaryonic flux-tubes by  $\rm{F_3}$ and  $\rm{F_{ \bar{3}}}$.
For the classified flux-tube sets, we can consider relationships between
them which are generated by quark pair creations and annihilations. The
division of a flux-tube is generated by a quark pair creation, and the
union of two flux-tubes by a pair annihilation. These relationships can
be used to construct topological spaces of flux-tubes, which are necessary
to define physical amplitudes. The assumptions for the construction of
topological spaces are
\begin{tabbing}
~~~\= (1) Open sets are stable flux-tubes.\\
 \> (2) The union of stable flux-tubes becomes a stable flux-tube.\\
 \> (3) The intersection between a connected stable flux-tube and \\
 \>~~~~ disconnected stable flux-tubes is the reverse operation of the union.
 
\end{tabbing} 
With these assumptions, we can follow the flux-tube sets that can be
produced from a given flux-tube by repeating union and intersection
operations. If the produced sets are closed under these operations, we
can classify the constructed topological spaces and this classification
procedure can be done by counting the numbers of incoming and outgoing
3-junctions in a given closed set. When we include the excited flux-tube
set $\rm{F_0}$, two 3-junctions can be created making it impossible to
assign fixed numbers of 3-junctions to a given set, and therefore,
we omit this possibility in this paper. Then the simplest non-trivial
topological space is
\begin{equation} 
\rm{T_0} = \rm{\{ \phi, ~ F_{1, \bar{1}}, ~ F^{2}_{1, \bar{1}},~ \cdots,~ F^{n}_{1, \bar{1}},~ \cdots }\} 
\end{equation} 
where $\rm{F^{n}_{1, \bar{1}}}$ represents n quarkonium meson states. Since $\rm{F_{1,\bar{1}}}$ can
be multiplied repeatedly without violating the law of baryon number
conservation, we may reduce the notation into $\rm{T_0 \equiv \{ \phi,~ F_{1, \bar{1}}\}}$. In
this notation, the topological space for baryon-meson system can be
represented as
\begin{equation} \rm{ T_1 = \{ \phi,~ F_3 \}}, 
\end{equation} 
and the baryon-meson-baryon space becomes
\begin{equation} \rm{ T_2 = \{\phi,~ F^{2}_{3} \}}. 
\end{equation} 
When outgoing 3-junctions exist, we need another index to represent the
topological space. For example, the space with two incoming 3-junctions
and one outgoing 3-junction is represented as
\begin{equation}\rm{T_{2, \bar{1}} = \{\phi,~ F^{2}_{3} F_{\bar{3}},~ F_3 F_{2, \bar{2}},~ F_{4, \bar{1}} \} } 
\end{equation} 
In general, we can write down the spaces as
\begin{equation}\rm{ T_{i, \bar{j}} = \{ \phi,~ F^{i}_{3} F^{j}_{\bar{3}},~ F^{i-1}_{3} F^{j-1}_{\bar{3}} F_{2, \bar{2}},~ \cdots \}} 
\end{equation}
where i is the number of incoming 3-junctions and j that of 
outgoing 3-junctions.\\
~ Now let's try to define physical amplitude related to the measures
on flux-tubes. It is physically natural to define the amplitudes A
for a quark to be connected to another quark or antiquark through
given flux-tube open set. In order to quantify A, we can assume
the existence of a measure M of A satisfying the conditions
\begin{tabbing}
~~${\rm (1)}~{ \rm  M(A)~ decreases~ as~ A ~increases},$\\
~~${\rm (2)}~\rm{ M(A_1)+M(A_2)~=~M(A_1 A_2)~ when~ A_1~ and~A_2~ are~independent}.$~~~
\end{tabbing} 
From these two conditions, the measure M of A can be solved as
functions of A
\begin{equation} 
\rm{M(A)} = {\rm -k} \ln {\frac{\rm A}{\rm A_0}}, 
\end{equation} 
where $\rm{A_0}$ is a normalization constant and k is an appropriate
parameter. One reasonable method to convert the amplitude A into
a concrete form is to consider the measure M as a metric
function defined on the flux-tube. A general form of distance function
between the two boundary points \rm{\bf{x}} and \rm{\bf{y}} can be written down as
$ | \rm{\bf{x}} - \rm{\bf{y}} |^\nu$ with $ \nu$ being an arbitrary number. This distance function
can be made metric for the points  \rm{\bf{z}} satisfying
\begin{equation} 
| \rm{\bf{x}} - \rm{\bf{z}} |^\nu + | \rm{\bf{z}} - \rm{\bf{y}} |^\nu \stackrel{>}{=} |\rm{\bf{x}} - \rm{\bf{y}} |^\nu . 
\end{equation} 
The set of points \rm{\bf{z}} not satisfying this triangle inequality can
be taken as forming the inner part of the flux-tube where it
is impossible to define a metric from boundary points with given $\nu$.
If we take $ | \rm{\bf{x}} - \rm{\bf{y}} |^\nu$ as an appropriate measure for A, we need
to sum over contributions from different $\nu'\rm{s}$. For a small increment
$\rm{d}\nu$, the product of the two probability amplitudes for $ |\rm{\bf{x}} - \rm{\bf{y}} |^\nu$ and
$ |\rm{\bf{x}} - \rm{\bf{y}}|^{\nu + d\nu}$ to satisfy the metric conditions can be accepted as the
probability amplitude for the increased region to be added to the inner
connected region which is out of the metric condition. Considering all
possibilities, the full connection amplitude becomes
\begin{equation} 
\rm{A = A_0 exp\{ -\frac{1}{k} \int^\alpha_1 F(\nu) r^\nu d\nu\}}, 
\end{equation} 
where the lower limit of $\nu$ is fixed to 1 because there exists
no point  \rm{\bf{z}} satisfying the triangle inequality with $ \nu < 1$, and
the upper limit $\alpha$ is arbitrary. The weight factor $\rm{F(\nu)}$ has
been introduced in order to account for possible different contributions
from different  $\nu'\rm{s}$, and the variable r is
\begin{equation} 
\rm{r} = \frac{1}{\it{l}} |\rm{\bf{x}} - \rm{\bf{y}} | 
\end{equation} 
with $l$ being a scale parameter. When we take the case of $ \alpha = 2$,
which corresponds to a spherical shape flux-tube, and the case of
equal weight  $\rm{F(\nu)} = 1$, we get
\begin{equation} 
\rm{A} = \rm{A_0}~ exp \{ -\frac{1}{k} 
~\frac{\rm{r^2} - \rm{r}}{\ln{\rm{r}} } 
\}. 
\end{equation}
We will use this form of connection amplitude.\\

 For scattering states, we need to formulate the connection
amplitude in momentum space. By applying the same arguments, we can
write down the connection amplitude for two boundary points with
momenta $\rm{\bf{p_1}}$ and $\rm{\bf{p_2}}$
\begin{equation} 
\rm{A} = \rm{A_0}~ exp \{ 
-\frac{1}{\tau} \int^{\alpha}_{1} G(\nu) |\rm{\bf{p_1}} - \rm{\bf{p_2}} |^{\nu} d\nu \}, 
\end{equation}
where $\rm{G(\nu)}$ and $\alpha$ are weight factor and the upper limit of
$\nu$. In case of $\rm{G(\nu)} = 1$ and $\alpha = 2$, we get
\begin{equation} 
\rm{A(p)} =A_0~ exp\{ 
-\frac{1}{\tau} 
~ \frac{\rm{p^2} - \rm{p}}{\ln{\rm{p}}} 
\} 
\end{equation} 
with $ \rm{p} = | \rm{\bf{p_1}} - \rm{\bf{p_2}} |$.\\

 Now let's consider the two partons with momenta $\rm{p_1}$ and $\rm{p_2}$ subtending
an angle $\theta$. We may take the parton 1 as a quark and the second
one as a radiated gluon, or may take both particles as quarks or
gluons. In any case, we want to calculate the probability amplitude to
have the third gluon with given momentum in some direction. In our
momentum space flux-tube model, the probability amplitude to have the
third gluon is taken to be proportional to the connection amplitude
representing the connections of third gluon with the other two particles.
Of course, these connections are formulated in non-perturbative region,
that is, in long range region. Our calculations have been carried out
with the form of A given in Eq.(13), varying the magnitude and angle
of the momentum $\rm{\bf{p}_3}$ of the third gluon. The amplitude is
\begin{equation} 
\rm{A} = \rm{A^2_0}~ exp\{ 
-\frac{1}{\tau} 
\frac{|\rm{\bf{p}_1} - \rm{\bf{p}_3}|^2 - |\rm{\bf{p}_1} - \rm{\bf{p}_3}|} 
{\ln{|\rm{\bf{p}_1} - \rm{\bf{p_3}}|}} 
-\frac{1}{\tau} 
\frac{|\rm{\bf{p}_3} - \rm{\bf{p}_2} |^2 - |\rm{\bf{p}_3} - \rm{\bf{p}_2}|} 
{\ln{|\rm{\bf{p}_3} - \rm{\bf{p}_2}|}} 
\}. 
\end{equation} 
 In Fig.1, the probability amplitudes to have the third gluon in the
same plane of $\rm{\bf{p}_1}$ and $\rm{\bf{p}_2}$ are shown as functions of the magnitude of
gluon momentum. We have fixed some parameters as $\rm{A_0} = 1$, $\tau = 0.7$ and
$\rm{p_1} = 1.0$, and we have shown the case of $\rm{p_2} = 0.3$ and $\theta =\frac{\pi}{6}$ in Fig.1.
The different curves correspond to different angles which have been
selected by dividing the given angle $\theta$ into 9 equally spaced angles.
The lower curve corresponds to smaller angle, and so we can see that
the maximum probability occurrs at higher values of momentum as the
value of angle increases. In any case, the maximum probability results
from a momentum value between the two given values $\rm{p_1} = 1.0$ and $\rm{p_2} = 0.3$.
We have also changed the values of $\rm{p_2}$ and $\theta$, and obtained similar
results. The general shape is closely related to the momentum or
energy distributions of particles in a jet\cite{jet}.\\

 For angular variations, we have checked the amplitude dependence on
angles by changing the values of $\rm{p_2}$ and $\theta$. For a fixed value of $\rm{p_2}$,
the shapes of the curves are not so much changed as we vary the
values of subtending angle $\theta$. However, if we vary the values of $\rm{p_2}$
for fixed $\theta$ as in Fig.2, we can see that the probability amplitude
becomes peaked around the direction of larger momentum. In Fig.2,
we have fixed $\theta$ to $\frac{\pi}{3}$ and the angle is measured from the axis of
larger momentum $\rm{p_1}$. We have obtained quite similar results for other
choices of $\theta$. For larger values of  $\rm{p_2}$ comparable with $\rm{p_1}$, the peak
values of A appear near $\frac{\theta}{2}$, and move toward the direction of
$\rm{p_1}$ as the value of $\rm{p_2}$ decreases. This result can be accepted as
consistent with the angular ordering condition in Eq(1). Moreover, we
can predict the maximal direction in which a new gluon will be radiated.
Since the variations of the probability amplitude are continuous, it will
be possible to introduce smooth function instead of step function $\Theta$ to
predict gluon radiation. In this case, the probability to violate the strict
angular ordering represented by  $\Theta(\theta - \theta')$ is not zero as can be seen in Fig.2.\\

 In summary, we have calculated the probability amplitude of gluon
radiation by using momentum space flux-tube model. We have found that
the direction of a radiated gluon depends on the magnitude and
directions of momenta of the two precedent partons. The new gluon will
be most likely radiated in the direction between the two initial partons,
which corresponds to the usual process of fragmentation\cite{frag}. This result is
consistent with the angular ordering assumption appearing in the equations
such as CCFM equation. However, there exists the possibility in our
formalism that the new gluon is to be radiated in a direction outside
the angle subtended by the two initial partons. We need further work
to replace the simple step function representing angular ordering with
some smooth function which can be managable in solving the equations
describing gluonic behaviors.\\

 One of us (Jong B. Choi) thanks the members of JLC Group in 
KEK for their hospitality during his stay, when the final form of
this paper has been accomplished.

\newpage		
\section*{Figure Caption}
\begin{itemize}
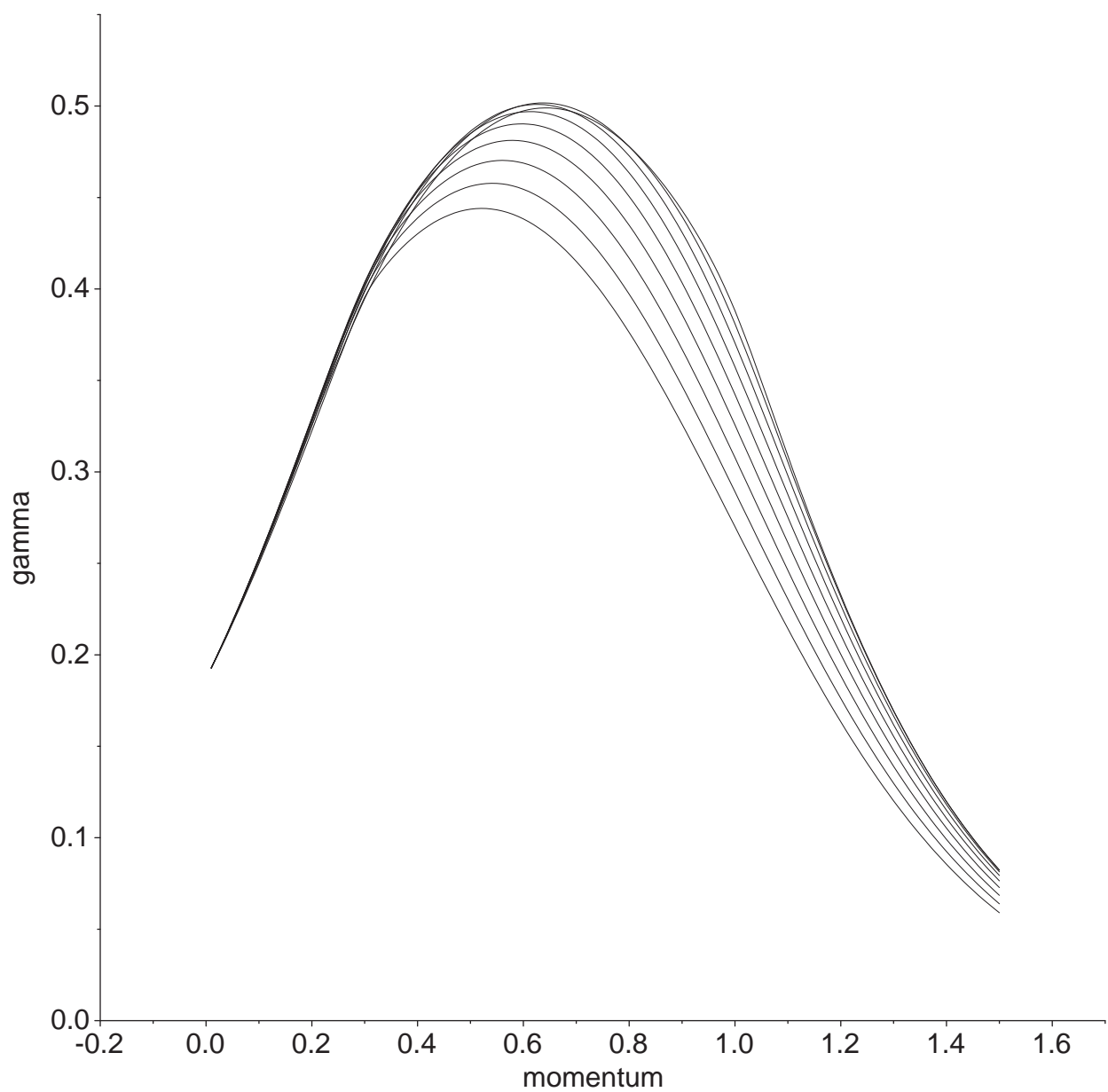
\item[Fig.~1]~~The probability amplitude to have new gluon as functions of gluon momentum. 
				We have fixed $\rm{p_1}=1.0, \rm{p_2}=0.3$ with the angle $\theta = \frac{\pi}{6}$.
				The different curves correspond to different directions.
				
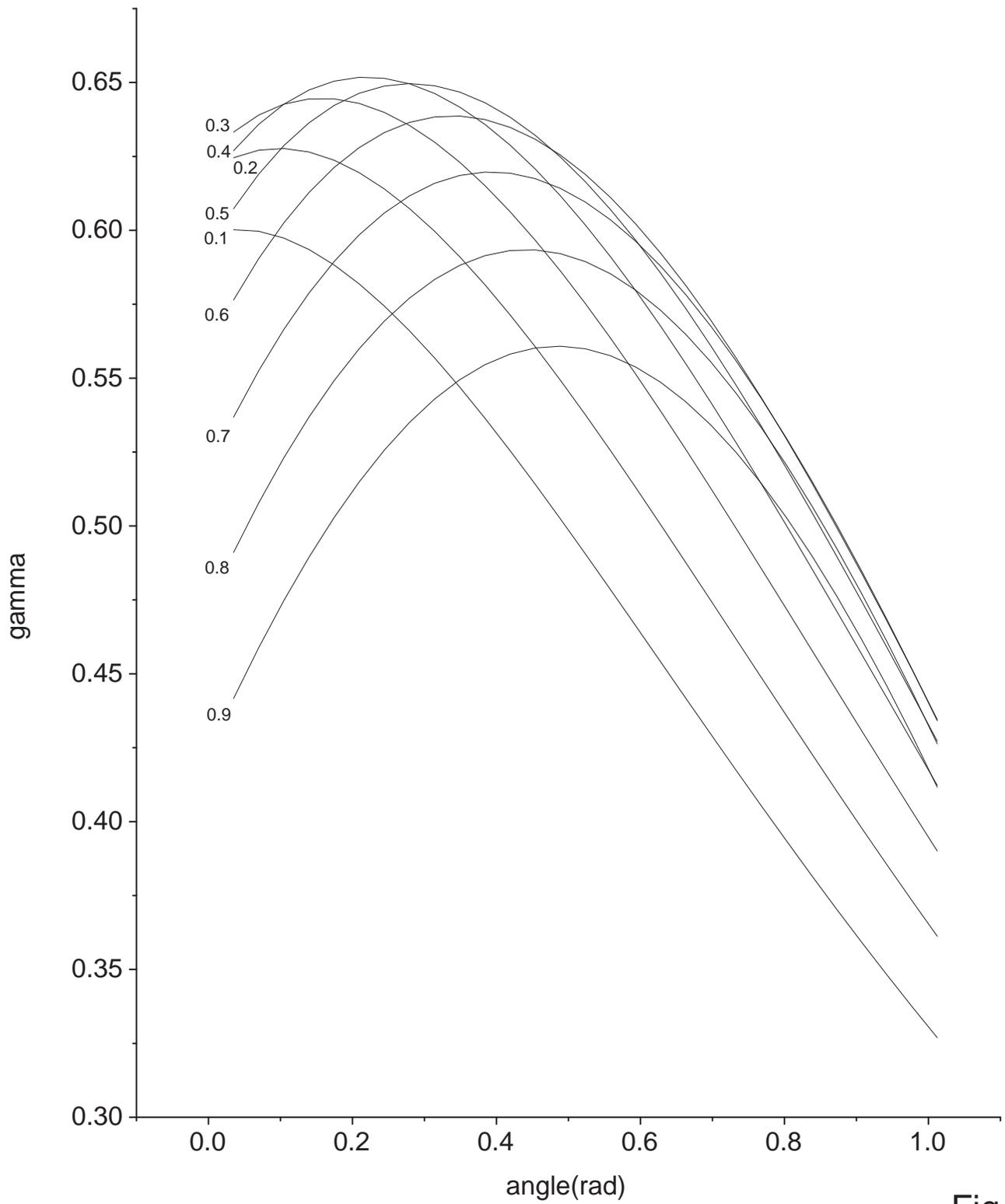
\item[Fig.~2]~~Angular variation of the probability amplitude. We fixed the angle 
				$\theta$ as $\frac{\pi}{3}$ and $\rm{p_1} = 1.0$. The different curves correspond to 
				different values of $\rm{p_2}$. 
\end{itemize}
\end{document}